\documentclass[reprint, amsmath,amssymb,aps,pra,]{revtex4-1}
\usepackage{color}
\usepackage{graphicx}
\usepackage{dcolumn}
\usepackage{bm}
\usepackage{amssymb}

\begin{document}

\preprint{APS/123-QED}

\title{Noncommutative field with constant background fields and neutral fermion}

\author{Cui-bai Luo$^{1}$}
\author{Feng-yao Hou$^{2,5}$}
\author{Zhu-fang Cui$^{3,5}$}
\author{Xiao-jun Liu$^{1,}$}\email{Email: liuxiaojun@nju.edu.cn}
\author{Hong-shi Zong$^{3,4,5,}$}\email{Email: zonghs@nju.edu.cn}

\address{$^{1}$ Key Laboratory of Modern Acoustics, MOE, Institute of Acoustics, and Department of Physics, Nanjing University, Nanjing 210093, China}
\address{$^{2}$ Institute of Theoretical Physics, CAS, Beijing 100190, China}
\address{$^{3}$ Department of Physics, Nanjing University, Nanjing 210093, China}
\address{$^{4}$ Joint Center for Particle, Nuclear Physics and Cosmology, Nanjing 210093, China}
\address{$^{5}$ State Key Laboratory of Theoretical Physics, Institute of Theoretical Physics, CAS, Beijing 100190, China}

\date{\today}

\begin{abstract}
Introducing constant background fields into the noncommutative gauge theory, we first obtain a Hermitian fermion Lagrangian which involves a Lorentz violation term, then we generalize it to a new deformed canonical noncommutation relations for fermion field. Massless neutrino oscillation in the deformed canonical noncommutation relations is analyzed. The restriction of the noncommutative coefficients is also discussed. By comparing with the existing experimental data of conventional neutrino oscillations, the order of noncommutative deformed coefficients is given from different ways.

\bigskip


\end{abstract}

\maketitle

\section{\label{sec:level1}introduction}
Neutrinos are massless particles in the Standard Model of particle physics. In order to explain neutrino oscillation, the conventional scenario is to assume that neutrino physics is beyond the Standard Model, and then get neutrinos with nonzero mass \cite{Minkowski77,yanagid79,Rnmohapatra1980}, where is a spectrum of three or more neutrino mass eigenstates, and flavor state is a mixing one of mass eigenstates \cite{Zmaki1962}. D. Colladay, V. A. Kosteleck\'y \cite{PhysRevD.55.6760}, S. Coleman and S. L. Glashow \cite{Coleman1997249} noted that neutrino oscillations may take place for massless neutrinos in which case that Lorentz invariance is violated in the neutrino sector.

Lorentz invariance violation could give a preparative solution to two important experimental problems, namely, the observation of TeV photons and of cosmic ray events above the GZK cutoff \cite{Coleman1997249,PhysRevD116008,Tkifune1999,Rjprotheroe2000,Gamelino2001,Wolfgang2011}, since the threshold energy at which the cutoff occurs could be altered by modifying special relativity. The high energy tests of Lorentz-violating depended on a perturbative framework for neutrinos was discussed in Ref.~\cite{PhysRevD116008}. The consequences of Lorentz and CPT violations to the three-generation neutrino oscillations were abundantly analyzed in the massless neutrino sector~\cite{barger2000, irina2002, degouv2002}. The Lorentz-violating extension of the minimal Standard Model~\cite{Dcolladay1998} including CPT-even and CPT-odd terms were also studied,
the stability and causality are investigated in quantum field theories that incorporating Lorentz and CPT violations \cite{Valankostelecky2001}.

The violation of Lorentz invariance may arise from quantum gravity \cite{jorge2002,PhysRevLett89,Mavromatos20041}, random dynamics \cite{froggatt2002}, string theory \cite{nikolao2002} and field theories with gravity \cite{Vakostelecky1989}. Another approaches is noncommutative field theory \cite{Lett141601, Guralnik2001450}, where any realistic noncommutative theory is found to be physically equivalent to a subset of a general Lorentz-violating standard-model extension involving ordinary fields. Spontaneous violation of CPT/Lorentz symmetries in string theories is also well known \cite{Mrdouglas2001}, where Lorentz symmetry breakdown is natural when the perturbative string vacuum is unstable.

In order to get the oscillation of massless neutrinos, the introduction of Lorentz- or (and) CPT-violating items into the Lagrangian is needed, but for this introduction, people often introduce directly from the Lorentz-violating extension of the Standard Model, instead of introducing directly by noncommutative fields theory. Different from the conventional way, J. M. Carmona \emph{et al} \cite{Jmcarmona2003} consider the Hamiltonian of free complex bosonic field theory and investigate the connection between Lorentz invariance violation and noncommutativity of fields by proposing a new Moyal Product between functionals, which is consistent with the commutation relations. In that case, they only discuss the effects of neutrinos on noncommutative space.

In this paper, since all noncommutative effects vanish for neutral fermions in noncommutative gauge theories, we introduce neither new Moyal Product nor Lorentz-violation extension of the Standard Model, but mainly focus on the constant background fields in noncommutative gauge theories, and obtain a Hermitian fermion Lagrangian which involves a Lorentz violation term. The Lorentz violation term deform the conventional canonical commutation relations. From this we try to generalize our model to a new deformed canonical noncommutation relations for fermion field. On this basis of the new canonical noncommutativity, we obtain massless neutrino Lagrangian which satisfies the deformed canonical relations. Subsequently, we investigate the restriction of the noncommutative coefficients. Comparing with the existing experimental data on conventional neutrino oscillations, finally the order of noncommutative deformation coefficients is given by different ways.

\section{\label{sec:level2}Noncommutative coefficients of neutral fermions with constant background fields}
The idea that the spacetime coordinates do not commute is quite old in mathematics and physics \cite{Hssnyder1947}. Noncommutativity \cite{Ihab2000, chaichian2003,chaichian2004, PhysRevLett.94.151602, SheikhJabbari201163} is the central mathematical concept expressing uncertainty in quantum mechanics, where it applies to any pair of conjugate variables, such as position and momentum. A. Connes \cite{Aconnes1985} \emph{et al} build a Noncommutative Geometry system, and tried to apply it to physics or even the Standard Model. N. Seiberg and E. Witten \cite{Nseiberg1999} extend earlier ideas about the appearance of noncommutative geometry in string theory with a nonzero $B$ field, and get an equivalence between ordinary gauge fields and noncommutative gauge fields, which is realized by a change of variables. The spacetime noncommutativity undergoes a revival following the realization which occurs naturally in string theory \cite{Aconnes1998}.

In noncommutativity framework, the canonical commutator of the coordinates in the spacetime manifold is
\begin{eqnarray} \label{eq:nc1}
[x^\mu, x^\nu] = i\theta^{\mu \nu} \label{eq:nc1}, \quad \quad [\partial_i, \partial_j] = 0,
\end{eqnarray}
with a parameter $\theta$ which is a real and antisymmetric (constant) tensor. In addition, there are operator and Lie Algebra forms
\begin{eqnarray} \label{eq:nc2}
[x^\mu, x^\nu] = i \hat \theta^{\mu \nu},    \quad \quad  [x^\mu, x^\nu] = i C^{\mu \nu}_{k} x^k.      \label{eq:nc2}
\end{eqnarray}
In the case of Eq. (\ref{eq:nc1}), a simple set of derivatives $\partial_i$ can be defined by the relations
\begin{eqnarray}  \label{eq:nc3}
\partial_i x^j =\delta^j_i, \quad \quad  [\partial_i, \partial_j] = 0
\end{eqnarray}
and the Leibnitz rule.

To construct a noncommutative quantum field theory, one approach is to replace conventional fields with noncommutative fields, and conventional products with Moyal Star Products \cite{Dbfairlie1964}, namely,
\begin{eqnarray} \label{eq:nc4}
f(x) \star g(x) = \mathrm{exp}(\frac{1}{2}i\theta^{\mu \nu} \partial_{x^\mu} \partial_{y^{\nu}})f(x)g(x) \vert_{x=y}.
\end{eqnarray}
It is well known that as noncommutative phase space one can occasionally generalize Eq. (\ref{eq:nc3}) to
\begin{eqnarray} \label{eq:nc5}
[x^\mu, x^\nu] = i\theta^{\mu \nu}, \quad [\partial_\mu, \partial_\nu] = -i\Phi_{\mu \nu}
\end{eqnarray}
to incorporate an additional background field \cite{Mrdouglas2001}.

Gauge theories, such as quantum electrodynamics (QED), is ordinarily transformed into noncommutative QED \cite{Ihab2000} by Moyal Star Products. Action of pure noncommutative $U(1)$ Yang-Mills and matter field are respectively \cite{Hayakawa2000394}
\begin{eqnarray} \label{eq:nca5}
&&S_{YM} = \int d^4x(- \frac{1}{4\mathit{q}^2})\hat{F}_{\mu \nu}\star \hat{F}^{\mu \nu}, \nonumber \\
&&S_{Matter}=\int d^4x (\frac{1}{2}i \bar {\hat{\psi}} \gamma^\mu \star \overleftrightarrow{\hat{D}}_\mu \hat{\psi} - m\bar {\hat{\psi}} \star \hat{\psi}) \nonumber
\end{eqnarray}
where $q$ is the coupling constant. Then for noncommutative QED, the Hermitian Lagrangian can be written as
\begin{eqnarray} \label{eq:nc6}
\mathcal{L} =\frac{1}{2}i \bar {\hat{\psi}} \gamma^\mu \star \overleftrightarrow{\hat{D}}_\mu \hat{\psi} - m\bar {\hat{\psi}} \star \hat{\psi} - \frac{1}{4q^2}\hat{F}_{\mu \nu}\star \hat{F}^{\mu \nu},
\end{eqnarray}
where $\hat A_\mu$ and $\hat \psi$ are the fields in noncommutative QED space, and $A_\mu$ together with $\psi$ are for conventional QED, and have
\begin{eqnarray} \label{eq:nc7}
&&\hat{F}_{\mu \nu} =\partial_\mu \hat{A}_\nu - \partial_\nu \hat{A}_\mu -i[\hat{A}_\mu,\hat{A}_\nu]\star,  \nonumber \\
&&\hat{D}_\mu \hat{\psi}=\partial_\mu \hat{\psi} -i\hat{A}_\mu \star \hat{\psi}, \nonumber \\
&&\hat{A} \star \overleftrightarrow{\hat{D}}_\mu \hat{B}= \hat{A} \star \hat{D}_\mu \hat{B} - \hat{B} \star \hat{D}_\mu \hat{A}.
\end{eqnarray}

Noncommutative coefficient $\theta^{\mu \nu}$ are acceptable only when $\theta^{0k} =0,~ \theta^{ij} \ne 0 ~(i,j =1,2,3)$, on account of the difficulties with perturbative unitarity \cite{Jgomis2000}. Nonetheless, presumably certain cases with $\theta^{0j} \ne 0$ can also be allowed, since the presence of observer Lorentz invariance implies that there is no difficulties with perturbative unitarity provided $\theta_{\mu \nu} \theta^{\mu \nu} >0, ~\theta_{\mu \nu} \tilde{\theta}^{\mu \nu} = 0$, which can be converted into one with only $\theta^{jk}$ nonzero by a suitable observer Lorentz transformation. It is similar to the condition applying for open bosonic stings where the presence of a nonzero $B^{jk}$ field is equivalent to a constant magnetic field on a $D_p$ brane \cite{Nseiberg1999}.

{N. Seiberg and E. Witten \cite{Nseiberg1999} proposed an map from the noncommutative $U(N)$ gauge theory to an ordinary $U(N)$ gauge theory. In 2002, Chaichian, Pre$\check{s}$najder \emph{et al} \cite{Chaichian2002132} present the No-Go Theorem, this theorem show that matter fields in the noncommutative $U(1)$ gauge theory can only have $\pm1$ or $0$ charges and for a generic noncommutative
$\Pi^n_{i=1}U(N_i)$ gauge theory matter fields can be charged under at most two of the $U(N_i)$ gauge group factors. In Refs. \cite{madore20001,
jurco2000, jurco2001, calmet2002}, the authors argued that the Seiberg–Witten map, which relates a noncommutative gauge theory to an ordinary one, paves the way for constructing the noncommutative version of gauge theories based on generic Lie algebras with matter fields in generic representations, thus circumventing the restrictions discussed in Ref. \cite{Chaichian2002132}. However, Just as shown in Ref. \cite{Chaichian201055} that the Seiberg-Witten map can only be consistently defined and used for the gauge theories which respect the no-go theorem.}

To the first order, the Seiberg-Witten map reads \cite{Nseiberg1999, jurcol2000, Jurco2001148, aabichl2002}
\begin{eqnarray} \label{eq:nc8}
\hat A_\mu &=& A_\mu -\frac{1}{2}\theta^{\gamma \rho}A_\gamma(\partial_\rho A_\mu+F_{\rho \mu})  \nonumber  \\
\hat \psi &=& \psi - \frac{1}{2}\theta^{\gamma \rho} \partial_\rho \psi.
\end{eqnarray}
Substitute Eq. (\ref{eq:nc4}) and Seiberg-Witten Map Eq. (\ref{eq:nc8}) into the Lagrangian (\ref{eq:nc6}), then in the new Lagrangian the gauge invariant is given manifestly, and it consists of ordinary QED plus nonrenormalizable Lorentz-violating corrections. It should be noted that noncommutative effects vanish for neutral fermions. One will get the same results if one use the Moyal Star Products (Eq. (\ref{eq:nc4})) in the case of canonical commutator of the coordinates Eq.(\ref{eq:nc1}), because of such a property of the Moyal Star Products
\begin{eqnarray} \label{eq:nc9}
\int \!\!  d^4x f(x)  \!\star  \!g(x)  \!\! =  \!\! \! \int  \!\!   d^4x g(x)\star f(x)  \!\! =   \! \!\!  \int  \!\!   d^4x f(x)  g(x).
\end{eqnarray}

{In the No-Go Theorem, neutral fermions could be in `adjoint' representation of noncommutative $U(1)$ gauge theory. The neutral fermions field, although is not carrying any $U(1)$ charge, similarly to noncommutative photons, carries the corresponding dipole moment \cite{Chaichian2002132}.} Considering the above situation, in order to get other noncommutative effects of neutral fermions, it seems to be a reasonable consideration by generalizing it's Lagrangian to noncommutative phase space (depending on Eq. (\ref{eq:nc5})), which is equivalent to attach a constant background field. But we will encounter some complex issues.

In this paper, in order to avoid these difficulties above, we mainly pay attention to this cases with constant electromagnetic background fields in noncommutative fields, and try to generalize this model to new canonical noncommutation relations. In the new noncommutativity, we assume that all fermion
be applied to this new ralations, then we investigate neutrino Lagrangian which involves a Lorentz violations term.

One can get the relation (\ref{eq:nc6}) when consider noncommutative QED under the canonical commutation relations Eq. (\ref{eq:nc1}), and then substitute Seiberg-Witten Map (Eq. (\ref{eq:nc8})) into the Lagrangian Eq. (\ref{eq:nc6}). The relation at the leading order in noncommutative coefficients is given as
\begin{eqnarray} \label{eq:nc10}
\mathcal{L} &=&\frac{1}{2}i \bar \psi \gamma^\mu \overleftrightarrow{D}_\mu \ {\psi} - m\bar \psi \psi  - \frac{1}{4} F_{\mu \nu}F^{\mu \nu} +\Theta(\theta^2) \nonumber \\
&+& \frac{1}{8}\theta^{\alpha \beta}q[2mF_{\alpha \beta}\bar \psi \psi +  F_{\alpha \beta}F_{\mu \nu} F^{\mu\nu}-4F_{\alpha \mu}F_{\beta \nu}F^{\mu\nu}] \nonumber \\
&+& \frac{1}{8}i\theta^{\alpha \beta}q[2F_{\alpha \mu} \bar \psi \gamma^\mu \overleftrightarrow{D}_\beta \ {\psi} - F_{\alpha \beta} \bar \psi \gamma^\mu \overleftrightarrow{D}_\mu \ {\psi}],
\end{eqnarray}
where we have redefined $A_{\mu} \to qA_{\mu}$. If we make the replacement
\begin{eqnarray} \label{eq:nc11}
F_{\alpha \beta} \to F_{\alpha \beta}+f_{\alpha \beta},
\end{eqnarray}
where $f_{\alpha \beta}$ is a constant background field and $F_{\alpha \beta}$ is understood to be a small fluctuation, it yields the Hermitian Lagrangian describing the leading-order effects of noncommutativity in constant background fields
\begin{eqnarray} \label{eq:nc12}
\mathcal{L} &=&\frac{1}{2}i \bar \psi \gamma^\mu \overleftrightarrow{D}^\nu (\mathrm{g}_{\mu \nu}+\eta_{\mu \nu}){\psi} - m\bar \psi \psi   \nonumber \\
&-& \frac{1}{4}(\mathrm{g}_{\alpha \mu} \mathrm{g}_{\beta \nu}+\kappa_{\alpha \beta \mu \nu})F^{\alpha \beta}F^{\mu \nu}+\Theta(\eta^2,\kappa^2),
\end{eqnarray}
where $\mathrm{g}_{\mu \nu}$ is the metric tensor, dimensionless coefficient $\eta_{\mu \nu}$ and $\kappa_{\alpha \beta \mu \nu}$ are constant tensors that associated with both the constant electromagnetic fields and the coefficients of canonical commutator,
\begin{eqnarray}\label{eq:nc13}
\eta_{\mu \nu}=-1/2 q f_{\mu}^{\phantom{1}\sigma}\theta_{\sigma \nu},
\end{eqnarray}
where $q$ is replaced with a scaled effective value $q\to (1+1/4 q f^{\mu \sigma}\theta_{\mu \sigma})q$.
If one remove the mass $m$, and let electromagnetic background fluctuation $F_{\mu \nu} \to 0$, then an effective Lagrangian is given
\begin{eqnarray} \label{eq:nc14}
\mathcal{L} &=&\frac{1}{2}i \mathrm{g}_{\mu \nu} \bar \psi \gamma^\mu \overleftrightarrow{\partial}^\nu  \mathrm \psi + \frac{1}{2}i \gamma^\mu\eta_{\mu \nu} \bar \psi  \overleftrightarrow{\partial}^\nu\psi.
\end{eqnarray}

Next, we generalize Eq. (\ref{eq:nc14}) to coupled field $\psi_{i,j}$
\begin{eqnarray} \label{eq:nc15}
\mathcal{L} &=& \frac{1}{2}i \bar \psi_{i} \gamma^\mu \overleftrightarrow{\partial_\mu} \psi_{i} +  \frac{1}{2}i \eta_{\mu \nu(ij)} \bar \psi_{i} \gamma^\mu  \overleftrightarrow{\partial^\nu} \psi_{j},
\end{eqnarray}
where $\eta_{\mu \nu(ij)}$ violate Lorentz invariance. Then we can obtain the conjugate momentum and the canonical commutation relations for fermion field,
\begin{eqnarray} \label{eq:nc16}
&&\pi_i= i(\delta_{ij}+\eta_{00(ij)})\psi^{\dagger}_{j}=\Lambda^{-1}_{ij}\psi^{\dagger}_{j}\nonumber \\
&&\{\psi_{i}(x), \psi^{\dagger}_{j}(y)\}=\Lambda_{ij}\delta(x-y),
\end{eqnarray}
with the coefficient of constant background field $\eta_{\mu \nu (ij)}$ diagonal in the spacetime indices, where the Lorentz violation term deform the conventional canonical commutation relations.

Next, let us put aside the background field, and assume that Eq. (\ref{eq:nc16}) applies to all fermion field. In fact, for fermion field, Eq. (\ref{eq:nc16}) inspired us to build new noncommutativity relations. We may assume a new noncommutativity by deforming the commutator of fields in analogy with the deformation commutator, such as Eq. (\ref{eq:nc16}). If it preserves the locality in the new set of canonical commutation relations, which become
\begin{eqnarray} \label{eq:nc17}
\{\pi_{\alpha}(x), \pi_{\beta}(y)\} &=& 0, \nonumber \\
\{\psi_{i}(x), \psi^{\dagger}_{j}(y)\} &=& \Lambda_{ij}\delta(x-y).
\end{eqnarray}

Now we can obtain an effective massless neutrino Lagrangian which satisfies this deformation noncommutative condition Eqs. (\ref{eq:nc16},\ref{eq:nc17}),
\begin{eqnarray} \label{eq:nc18}
\mathcal{L} = \frac{1}{2}i \bar \nu_{\alpha} \gamma^\mu \overleftrightarrow{\partial_\mu} \nu_{\alpha} +  \frac{1}{2}i \eta_{\mu \nu(\alpha \beta)} \bar \nu_{\alpha} \gamma^\mu  \overleftrightarrow{\partial^\nu} \nu_{\beta}，
\end{eqnarray}
where $\alpha, \beta$ are flavor indices, $\mu, \nu$ are spinor indices, and constant tensor $\eta_{\mu \nu (\alpha \beta)}$ is related to the flavor and spinor indices. There it contains renormalizable Lorentz-viollating corrections. On the basis, the Lagrangian may provide a general rotation-invariant model of three active massless neutrinos in noncommutative field theory.

The coefficients $\eta$ on deformation noncommutation relations violates Lorentz invariance but not CPT symmetry which is in agreement with Refs. \cite{PhysRevLett5265,brustein2002}. A general form for the quadratic sector of a renormalizable Lorentz and CPT-violating Lagrangian that describing a single massive spin-$\frac{1}{2}$ is given in Ref. \cite{Valankostelecky2001}
\begin{eqnarray} \label{eq:nc19}
\mathcal{L} =\frac{1}{2}i \bar \psi \Gamma^\mu \overleftrightarrow{\partial}_\mu \psi -   \bar \psi M \psi,
\end{eqnarray}
where
\begin{eqnarray} \label{eq:nc20}
\Gamma^\mu &:=& \gamma^\mu+c^{\nu \mu}\gamma_\nu+d^{\nu \mu}\gamma_5 \gamma_\nu+e^\mu+i f^{\mu}\gamma_5+\frac{1}{2}g^{\lambda \nu \mu }\sigma_{\lambda \nu},       \nonumber \\
M &:=& m+a_\mu \gamma^\mu +b_\mu \gamma_5 \gamma^\mu +\frac{1}{2}\rho ^{\mu \nu} \sigma_{\mu \nu}.  \label{eq:lv2}
\end{eqnarray}
Coefficients for Lorentz violation are all real, $c_{\mu \nu}$ and $d_{\mu \nu}$ are traceless, $g_{\lambda \mu \nu}$ antisymmetric in its first two indices, and $\rho_{\mu \nu}$ antisymmetric. Just as it pointed out in Ref. \cite{Valankostelecky2001} that all the parameters violate Lorentz invariance, while $a_\mu, b_\mu, e_\mu, f_\mu, g_{\lambda \mu \nu}$ also break CPT. Under the above assumptions, these parameters about Lorentz and CPT violation are given when they are applied to massless neutrinos \cite{P69016005}. On this base, Ref. \cite{P69016005} discussed the general equations of motion for the free propagation of neutrinos, which can be written as a first-order differential operator acting on the object $\nu_{\beta}$.

Now we go back to Eq. (\ref{eq:nc18}), where the constant tensor $\eta_{\mu \nu}$ denotes the noncommutative effects, and it violates Lorentz invariance but keeps CPT symmetry. For the convenience of calculation, the Lagrangian can be transformed into another form since divergence terms have no contribution to the action
\begin{eqnarray} \label{eq:nc21}
\mathcal{L} &=& \frac{1}{2}i \bar \nu_{\alpha} \gamma^\mu \partial_\mu \nu_{\alpha} +  \frac{1}{2}i  \eta_{\mu \nu(\alpha \beta)} \bar \nu_{\alpha} \gamma^\mu \partial^\nu \nu_{\beta},
\end{eqnarray}
and the Euler Lagrange equation of motion satisfy
\begin{eqnarray} \label{eq:nc22}
&&i\gamma^0\partial_0 \nu_{\alpha} + i\gamma^i\partial_i \nu_{\alpha} +i \eta_{\mu \nu(\alpha \beta)}\gamma^{\mu}\partial^{\nu}\nu_{\beta}=0.
\end{eqnarray}

Comparing Eq. (\ref{eq:nc22}) with the conventional equations of motion $(i\delta_{\alpha \beta}\partial_0 -\mathcal{H}_{\alpha \beta})\nu_\alpha =0$, the Hamiltonian for neutrinos is given as
\begin{eqnarray}\label{eq:nc23}
\mathcal{H}_{\alpha \beta} =-i\gamma^0\gamma^i\partial_i -i\eta_{\mu \nu(\alpha \beta)}\gamma^0\gamma^\mu \partial^\nu.
\end{eqnarray}

In this paper, there are four parameters: $\theta^{'\mu \nu}$ in the canonical commutator of the coordinates in the spacetime manifold (\ref{eq:nc1}),  $\theta^{\mu \nu}$ in the noncommutative field theory (\ref{eq:nc6}), $\eta^{\mu \nu}$ in the new deformed canonical commutation relations (\ref{eq:nc18}) and $\eta^{\mu \nu}$ in the Lorentz- and CPT-violating extension of the standard model (\ref{eq:nc19}), respectively. As the above analysis, in the canonical commutator of the coordinates, $\theta^{'\mu \nu}$ is a real and antisymmetric (constant) tensor, in the noncommutative field theory it meets this relationship $\theta^{\mu \nu}\theta_{\mu \nu} >0$ and $\theta^{\mu \nu} \tilde{\theta}_{\mu \nu} = 0$. The new neutrino Lagrangian and the new commutation relation (\ref{eq:nc18}) requiring hermiticity implies the noncommutative coefficients $\eta^{\mu \nu}$ are hermitian in generation space. Some features of this model are similar to the conventional massive neutrino case, however, there is unusual energy dependence. Moreover, for the single fermion extension model (\ref{eq:nc19}), the dimensionless coefficients $\eta^{00}>0$ and $\eta^{00}<0$ (which resembles the noncommutative coefficient $\eta_{\mu \nu}$) implies that instabilities and microcausality-violation arise at the Planck scale \cite{Valankostelecky2001}. However, in Eq. (\ref{eq:nc18}), we do not need to consider this condition. The restriction of rotation invariance provides an additional special limit of the theory Eq. (\ref{eq:nc18}), which can significantly reduce the complexity of calculations. In addition, restricting Hamiltonian to rotation-invariant leaves two coefficients $\eta^{00}_{\alpha \beta}$ and $ \eta^{ij}_{\alpha \beta} = \frac{1}{3} \eta^{kk}_{\alpha \beta} \delta^{ij}$. It may be assumed $\eta^{00}_{\alpha \beta}- \eta^{jj}_{\alpha \beta}=0$ since the trace component $\mathrm{g}_{\mu \nu} \eta^{\mu \nu}$ is Lorentz invariant and can be absorbed into the usual kinetic term (it is unobservable), so only one of two matrices with noncommutative coefficients is independent, leaves only the matrices $ \eta_{00(\alpha \beta)}$.

\section{\label{sec:level2}The order of noncommutative coefficient}
From what has been discussed above, the Hamiltonian of massless neutrino in deformation canonical commutation relations space is given, and the restriction to the noncommutative coefficients $\eta$ is also discussed. In the following sections, we will discuss the order of noncommutative coefficient by means of different methods (more detail can be found in Refs. \cite{P69016005,Arias2007401}).

With the relation (\ref{eq:nc18}), the Hamiltonian matrix for neutrinos can be given, substituting it into the Euler-Lagrange equation of motion, the general dynamical equation for neutrinos could be obtained as
\begin{eqnarray}\label{eq:24}
h_{\alpha \beta}&=&|\vec{p}|\delta_{\alpha \beta}
 \left(\begin{array}{cc}
1&0 \\
0&1
\end{array} \right)   \nonumber \\
&+& \frac{1}{|\vec{p}|}
\left(\begin{array}{cc}
-[\eta^{\mu \nu}p_{\mu}p_{\nu}]_{\alpha \beta}&0 \\
0&-[\eta^{\mu \nu}p_{\mu}p_{\nu}]^{\ast}_{\alpha \beta}
\end{array} \right).
\end{eqnarray}
The four momentum may be taken as $p_{\mu}=(|\vec{p}|,-\vec{p})$ at leading order.

For the convenience of the following discussions, we first consider the conventional neutrino oscillation. In the mechanism of conventional oscillation of massive neutrino, the simplest form of probability for oscillation between two species of particles with a mixing angle $i,j$ and energy levels $E_i, E_j$ is given as
\begin{eqnarray}\label{eq:25}
P_{i \to i}&=&1- \mathrm{sin}^2 (2 \theta)\mathrm{sin}^2 (\frac{1}{4E}\Delta m^2L),  \nonumber \\
P_{i \to j}&=&\mathrm{sin}^2 (2 \theta_{ij})\mathrm{sin}^2(\frac{1}{2}\Delta E_{ij}t)  \nonumber \\
&=&\mathrm{sin}^2 (2 \theta_{ij})\mathrm{sin}^2[\frac{1}{2}(\sqrt{p^2+m_{i}} -\sqrt{p^2+m_{j}})t]  \nonumber \\
&\approx&\mathrm{sin}^2 (2 \theta_{ij)}\mathrm{sin}^2 (\frac{1}{4E}\Delta m^2_{ij}L).
\end{eqnarray}

Then we turn back to the deformation canonical noncommutation relations (\ref{eq:nc16}) and the Hamiltonian density (\ref{eq:nc23}), the dynamical equation can be written as
\begin{eqnarray}\label{eq:26}
\dot{\nu}_{\alpha}=-\Lambda_{\alpha \beta}(\alpha \cdot \nabla \nu_{\beta}),
\end{eqnarray}
where $\alpha=\gamma^0 \vec{\gamma}$. Turning it into momentum space, the form above can be written as
\begin{eqnarray}\label{eq:27}
\mathit{E}\nu_{\alpha}=\Lambda_{\alpha \beta}(\alpha \cdot \vec{p} \!\quad \! \! \! \nu_{\beta}).
\end{eqnarray}
For simplicity, we choose $\Lambda$ matrix and unitary matrix $\Gamma$ in two dimensions
\begin{eqnarray}\label{eq:28}
\Lambda = \left(
\begin{array}{cc}
1& \lambda   \\
\lambda^{*}& 1
\end{array} \right), \qquad
\Gamma = \left(
\begin{array}{cc}
\tau & 1/\sqrt{2}  \\
-\tau& 1/\sqrt{2}
\end{array} \right),
\end{eqnarray}
where $\Gamma$ is diagonalized, $\Lambda$, $\lambda^*/\tau-\tau \lambda=0$ and $\lambda$ are associated with $\eta_{\alpha \beta}$. Then the energy spectrum for the neutrino is
\begin{eqnarray}\label{eq:29}
E^{\pm}_{i,j}=\pm (1+\varepsilon^{ij}\vert \lambda \vert)|\vec{p}|,
\end{eqnarray}
with $\varepsilon^{ij}$ the two-dimensional Levi-Civita symbol.

Taking $\nu_1$ and $\nu_2$ as the eigenstates of energy values in Eq. (\ref{eq:29}), the energy eigenstates can be determined through the diagonalized matrix $\Gamma$
\begin{eqnarray}\label{eq:30}
\left(\begin{array}{c}
\nu_{1} \\
\nu_{2}
\end{array} \right)=\Gamma \cdot \left( \begin{array}{c}
\nu_{\alpha} \\
\nu_{\beta}
\end{array} \right),
\end{eqnarray}
where $\nu_{1,2}$ and $\nu_{\alpha, \beta}$ represent different energy eigenstates and the flavor eigenstates. The time evolution can be determined by energy spectrum as follows
\begin{eqnarray}\label{eq:31}
\nu_{i,j}(t) &=& e^{-iE_{i,j}^{+}t+i\vec{p}\cdot \vec{x}}\nu_{1,2}(0) \nonumber \\
&=&\kappa\cdot(\varepsilon^{ij} \tau \cdot \nu_{\alpha}+ \nu_{\beta}),
\end{eqnarray}
with coefficients $\kappa$ relevant to unitary matrix coefficients $\tau$. The above equation can be parameterized as
\begin{eqnarray}\label{eq:32}
\nu_{1}=\mathrm{cos} \theta_{12}\nu_{\alpha}+\mathrm{sin}\theta_{12}\nu_{\beta},\nonumber \\
\nu_{2}=-\mathrm{sin} \theta_{12}\nu_{\alpha}+\mathrm{cos}\theta_{12}\nu_{\beta}.
\end{eqnarray}
Depending on relations (\ref{eq:30}), the above relations can be inverted, then the evolution of flavor state $\nu_{\alpha}$ in time $t$ can be given as
\begin{eqnarray}\label{eq:33}
\nu_{\alpha}(t)&=&[(\mathrm{cos}^2 \theta_{12}e^{-iE^+_{1}t}+\mathrm{sin}^2 \theta_{12}e^{-iE^+_{2}t)})\nu_{\alpha}(0)\nonumber \\
&+&\frac{1}{2}\mathrm{sin}2 \theta_{12}(e^{-iE^+_{1}t}-e^{-iE^+_{2}t})\nu_{\beta}(0)]e^{i\vec{p}\cdot \vec{x}}.
\end{eqnarray}
Then the probability of finding the flavor state $\nu_{\beta}$ at time $t$ can be given by
\begin{eqnarray}\label{eq:34}
P_{\nu_{\alpha} \to \nu_{\beta}}&=&\mathrm{sin}^2(2 \theta_{12})\mathrm{sin}^2 (|\vec{p}| |\lambda|t)  \nonumber \\
&\approx&\mathrm{sin}^2(2 \theta_{12})\mathrm{sin}^2( E |\lambda|L).
\end{eqnarray}
The above probability relations can be inverted by taking the fact that for $|\lambda| \ll 1, E\sim |\vec{p}|$, and velocities close to $c, t\to L$(the path length traversed by the neutrino).

The analysis above for two neutrino flavors can be extended to incorporate three flavors when generalize the deformation parameters and the mixing angles. Then the probability for oscillation between neutrino flavors can be given as follows
\begin{eqnarray}\label{eq:35}
P_{\nu_{\alpha} \to \nu_{\beta}} \approx  \mathrm{sin}^2(2 \theta_{ij})\mathrm{sin}^2( E |\lambda_{ij}|L),
\end{eqnarray}
where $i,j=1,2,3$. Comparing with the conventional massive neutrinos relations (\ref{eq:25}) and using the above results from the solar neutrino and atmospheric neutrino experiments with $\Delta m^2_{ij}$ and $\mathit{E}$, one can get relation $\Delta m^2_{ij}/4\mathit{E}^2=|\lambda_{ij}|$. The solar and atmospheric neutrino experimental data \cite{PhysRevD.78.033010, PhysRevLett.94.081802} was given as $\Delta m^2_{12} \sim 7.67 \times 10^{-5} \mathrm{eV}^2, E \sim 1\mathrm{MeV}$ and $\Delta m^2_{23} \sim 2.32 \times 10^{-3} \mathrm{eV}^2, E \sim 1.8 \mathrm{GeV}$, then the order of the deformation parameters is given as follows
\begin{eqnarray}\label{eq:36}
|\lambda_{12}| \sim 10^{-17},  \quad |\lambda_{23}| \sim 10^{-22}.
\end{eqnarray}

Works above only involve the two-generation special case, next, we turn to another useful parametrization of the noncommutative coefficient by CKM-like mixing angles and phases. Restricting $h_{\alpha \beta}$ to rotation-invariant models leaves only coefficients $\eta^{00}_{\alpha \beta}$ and $ \eta^{ij}_{\alpha \beta} = \frac{1}{3} \eta^{kk}_{\alpha \beta} \delta^{ij}$. Since the trace component $\mathrm{g}_{\mu \nu} \eta^{\mu \nu}$ is Lorentz invariant and can be absorbed into the usual kinetic term (it is unobservable), so it may be assumed as zero for convenience, only one of these matrices is independent. Taking into account the frame dependent of the theory latter, one assume rotation invariance in the Sun-centered frame $(S,x,y,z)$ for definiteness, then the effective Hamiltonian (\ref{eq:24}) reduces to the block-diagonal form
\begin{eqnarray}\label{eq:37}
(h_{sc})_{\alpha \beta} =\mathrm{diag}\lbrace (-\frac{4}{3}\eta^{SS}E)_{\alpha \beta}, (-\frac{4}{3}\eta^{SS}E)^{\ast}_{\alpha \beta}\rbrace,
\end{eqnarray}
where the irrelevant kinetic term is dropped.

The features of the Hamiltonian above are similar to the conventional massive neutrino case. It provides a general rotationally invariant model of three active neutrinos. Similar to the analysis of conventional massive neutrino mixing proceeds, the effective Hamiltonian can be diagonalized with a unitary matrix $U_{sc}$
\begin{eqnarray}\label{eq:38}
h_{sc} = U^{\dagger}_{sc}E_{sc} U_{sc},
\end{eqnarray}
where $E_{sc}$ is a diagonal matrix.

The deformation noncommutative coefficient matrix can be parameterized with three eigenvalues and a constant unitary matrix $A$. For each coefficient $\eta^{\mu \nu}$ has
 \begin{eqnarray}\label{eq:39}
 \eta^{\mu \nu} = (A^{\mu \nu})^{\dagger} \mathrm{diag}\{\eta^{\mu \nu}_{(1)}, \eta^{\mu \nu}_{(2)}, \eta^{\mu \nu}_{(3)}\} A^{\mu \nu}
 \end{eqnarray}
The unitary diagonalizing matrices $A^{\mu \nu}$ are given so that the Hamiltonian above takes the block diagonal form as
\begin{eqnarray}\label{eq:40}
U_{sc} =   \left(
\begin{array}{cc}
A&0   \\
0&A^{\ast}
\end{array} \right).
 \end{eqnarray}

The above decomposition is frame dependent as we have discussed, there one restrict this decomposition to the standard Sun-centered celestial equatorial frame as mentioned. In order to get a CKM-like decomposition of the $U$ matrices we take mixing angles and phases with $\eta^{\mu \nu}$ by $\beta^{\mu \nu}_{(12)}, \beta^{\mu \nu}_{(13)},\beta^{\mu \nu}_{(23)}$ and $\delta^{\mu \nu}, \gamma^{\mu \nu}_{1},\gamma^{\mu \nu}_{2}$. Then the $U$ matrices can be written explicitly. Here, we ignore the tedious steps,
\begin{eqnarray}\label{eq:41}
U^{\mu \nu}=  \Bigg(
CKM-like  \Bigg)
 \left(
\begin{array}{ccc}
1&0&0   \\
0&e^{i\gamma_1^{\mu \nu}}&0 \\
0&0&e^{i\gamma_2^{\mu \nu}}
\end{array} \right).
 \end{eqnarray}
In addition, considering that the $\gamma$ matrix of phases can be absorbed into the amplitudes in the conventional massive neutrino analysis, so these $\gamma$ phases can be neglected.

We replace $\eta^{\mu \nu}$ with $\eta^{SS}$ since we have assumed rotation invariance in the Sun-centered frame. To mimic the usual massive neutrino solution, there we consider only to taking vanishing phases, $\theta^{SS}_{13}$ and $\theta^{SS}_{23}=\pi/4$. This leaves three degrees of freedom, two eigenvalue differences and one mixing angle $\theta^{SS}_{12}$.

Using time evolution operator $S_{\alpha \beta}=  U^{\dagger}_{sc}e^{-iE_{sc}t} U_{sc}$, then the probabilities for massless neutrino of $\nu_{\beta}$ oscillating into neutrino of $\nu_{\alpha}$ in time $t$ can be written as $P_{\nu_{\beta} \to \nu_{\alpha}}=|S_{\nu_{\alpha} \nu_{\beta}}(t)|^2$. The probabilities are given as
\begin{eqnarray}\label{eq:42}
P_{\nu_{e} \to \nu_{\nu}} &=& P_{\nu_{e} \to \nu_{\tau}} =\frac{1}{2} \mathrm{sin}2 \theta \mathrm{sin}^2( \lambda \cdot EL/2) \nonumber \\
P_{\nu_{e} \to \nu_{e}} &=& 1 - \mathrm{sin}2 \theta \mathrm{sin}^2( \lambda \cdot EL/2),
\end{eqnarray}
where $\lambda = \frac{4}{3}[\eta^{SS}_{(2)} - \eta^{SS}_{(1)} ]$. Considering the LSND and KamLAND results $P_{\bar{\nu}_{e} \to \bar{\nu}_{e}} = 61\%,1 \mathrm{Mev} \le E\le 10 \mathrm{Mev}$ and $L=138 \sim 214 \mathrm{km}$, and $P_{\bar{\nu}_{e} \to \bar{\nu}_{e}} = 26\%, E\sim 45 \mathrm{Mev}$ and $L=30\mathrm{m}$, replacing $\Delta m^2/2 E$ with $\lambda E$ as the previous calculation methods , we get
\begin{eqnarray}\label{eq:43}
\lambda \sim 10^{-17}.
\end{eqnarray}

In addition, considering from this relation (\ref{eq:24}), one just study the left-handed neutrino and ignore the right-handed antineutrinos because the two terms have the similar form, then one have the simplified Hamiltonian as
\begin{eqnarray}\label{eq:44}
h=|\vec{p}|\delta_{\alpha \beta}+\eta_{\mu \nu (\alpha \beta)}\frac{p^{\mu}p^{\nu}}{|\vec{p}|},
\end{eqnarray}
where minus sign has been incorporated into the deformation noncommutative coefficients.

To reduce the deformation noncommutative coefficients, we assume the neutrino Hamiltonian matrix can be simplified and then diagonalized by unitary matrix $\Gamma$,
\begin{eqnarray}\label{eq:45}
 \Gamma^{\dagger} \mathcal{H} \Gamma&=& \Gamma  \left(\begin{array}{ccc}
E&\eta^{00}_{e\mu}\mathit{E}&0 \\
\eta^{00}_{e\mu}\mathit{E}&E+\eta^{00}_{\mu \mu}\mathit{E}&\eta^{00}_{\mu \tau}\mathit{E}\\
0& \eta^{00}_{\mu \tau}\mathit{E}&E
\end{array} \right) \Gamma^{\dagger} \nonumber \\
&=&E \left(\begin{array}{ccc}
1&0&0 \\
0&1+A-B&0\\
0&0&1+A+B
\end{array} \right),
\end{eqnarray}
where $\eta^{00}_{e\mu}$, $\eta^{00}_{\mu \tau}$, and $\eta^{00}_{ \mu \mu}$ are three nonzero noncommutative parameters in the specific model, the other noncommutative parameters are zero, and
\begin{eqnarray}\label{eq:46}
A= \eta^{00}_{\mu \mu}/2,\quad B=\sqrt{(\eta^{00}_{e\mu})^2+(\eta^{00}_{\mu \tau})^2+(\eta^{00}_{ \mu \mu})^2/4}.
\end{eqnarray}
Then the eigenstates can be the linear combination of flavor eigenstates since the unitary matrix $\Gamma$.

Using the same method mentioned above, with the unitary matrix $\Gamma$ and the diagonal eigenenergy matrix, finally the oscillation probabilities are given as
\begin{eqnarray}\label{eq:47}
P_{\nu_{e} \to \nu_{\mu}}&=&\frac{(\eta^{00}_{e \mu})^2}{B^2}\cdot \mathrm{sin^2}(BEL), \nonumber \\
P_{\nu_{\mu} \to \nu_{\tau}}&=&\frac{(\eta^{00}_{\mu \tau})^2}{B^2}\cdot \mathrm{sin^2}(BEL),\nonumber \\
P_{\nu_{\mu} \to \nu_{\mu}}&=&1-\frac{(\eta^{00}_{e \mu})^2+(\eta^{00}_{\mu \tau})^2}{B^2}\cdot \mathrm{sin^2}(BEL).
\end{eqnarray}

Comparing the above equation with the analysis of oscillation on noncommutative space such as mentioned in the previous content, then the order of noncommutatival parameters can be identified
\begin{eqnarray}\label{eq:48}
|\eta^{00}_{e\mu}|  \sim 10^{-17}, |\eta^{00}_{\mu \tau}|  \sim 10^{-22}.
\end{eqnarray}

Depending on the data above, the remaining coefficient $\eta^{00}_{\mu \mu}$ can be chosen to match other experimental data, the order of the remaining noncommutative parameters can be identified. Comparing in three relations (\ref{eq:36},\ref{eq:43},\ref{eq:48}), it is easy to find that one can get the same order of noncommutative coefficients by different methods which is expected in advance. Just as shown in this paper, the Lagrangian which involves the new deformation canonical noncommutation relations may give a model of three active massless neutrinos.

\section{\label{sec:level2}Summary}
In order to get the oscillation of massless neutrinos, the conventional scenario is depending on introducing Lorentz-violation extension of the Standard Model. Different from the previous treatment, in this paper we try to get oscillation of the massless neutrinos mainly by introducing new canonical anticommutation relations.  Considering that all noncommutative effects vanish for neutral fermions in noncommutative gauge theory, and the generalization of Lagrangian to noncommutative phase space will encounter some complex issues (such as the Moyal Star Products can not apply to this case), we directly introduce constant background fields into the noncommutative gauge theory and obtain a Hermitian fermion Lagrangian which involves a Lorentz violations term. The Lorentz violation term deform the conventional canonical commutation relations. Then we try to generalize this model to a new deformed canonical noncommutation relations for fermion field. On this basis of the new deformed canonical noncommutation relations, we investigate massless neutrino Lagrangian which satisfies the deformed canonical relations. The coefficient $\eta$ in massless neutrino Lagrangian violates Lorentz invariance but keep CPT symmetry, the restriction of the noncommutative coefficients is also discussed. By comparing with the existing experimental data of conventional neutrino oscillations, the order of noncommutative deformed coefficients is given. From the discussion above, we conclude that the Lagrangian which satisfy deformed canonical noncommutation relations may give a model of three active massless neutrinos.

\begin{acknowledgments}
This work is supported in part by the National Natural Science Foundation of China (under Grants Nos. 11275097, and 11475085), the Foundation of Graduate School of Nanjing University (under Grant No. 2014CL02), the National Basic Research Program of China (under Grant No. 2012CB921504), and the Jiangsu Planned Projects for Postdoctoral Research Funds (under Grant No. 1402006C).
\end{acknowledgments}

\bibliography{DW11443}

\end{document}